\documentclass[fleqn,10pt]{wlscirep}
\usepackage[utf8]{inputenc}
\usepackage[T1]{fontenc}
\usepackage{lineno}
\usepackage{multirow}
\usepackage{subcaption}

\title{DiaTrend: A dataset from advanced diabetes technology to enable development of novel analytic solutions}

\author[1,3,*]{Temiloluwa Prioleau}
\author[1]{Abigail Bartolome}
\author[2]{Richard Comi}
\author[3]{Catherine Stanger}

\affil[1]{Dartmouth College, Department of Computer Science, Hanover, 03755, USA}
\affil[2]{Dartmouth Health, Geisel School of Medicine, Lebanon, 03766, USA}
\affil[3]{Dartmouth College, Center for Technology and Behavioral Health, Lebanon, 03766, USA}

\affil[*]{corresponding author(s): Temiloluwa Prioleau (tprioleau@dartmouth.edu)}

\begin{abstract}
Objective digital data is scarce yet needed in many domains to enable research that can transform the standard of healthcare. While data from consumer-grade wearables and smartphones is more accessible, there is critical need for similar data from clinical-grade devices used by patients with a diagnosed condition. The prevalence of wearable medical devices in the diabetes domain sets the stage for unique research and development within this field and beyond. However, the scarcity of open-source datasets presents a major barrier to progress. To facilitate broader research on diabetes-relevant problems and accelerate development of robust computational solutions, we provide the DiaTrend dataset. The DiaTrend dataset is composed of intensive longitudinal data from wearable medical devices, including a total of 27,561 days of continuous glucose monitor data and 8,220 days of insulin pump data from 54 patients with diabetes. This dataset is useful for developing novel analytic solutions that can reduce the disease burden for people living with diabetes and increase knowledge on chronic condition management in outpatient settings.

\end{abstract}

\begin{document}

\flushbottom
\maketitle

\thispagestyle{empty}


\section*{Background \& Summary}

Advanced technologies like continuous glucose monitors (CGMs) and insulin pumps are transforming the standard of care for diabetes management \cite{beck2019advances,american20217,cappon2019continuous}. The ubiquitous nature of these devices enables real-time monitoring and treatment in daily living; this is a huge advantage over single point-in-time alternatives like glucose meters and insulin pens. Research shows that many patients with diabetes achieve better outcomes with CGMs and insulin pumps \cite{rodbard2017continuous,taylor2018effectiveness}. However, research also shows that digital data from these devices is significantly underutilized to optimize outcomes \cite{bartolome2021glucomine,bartolome2022computational}. Meanwhile, the next generation of solutions needed to advance diabetes care, such as the hybrid and fully closed-loop artificial pancreas \cite{thabit2016coming,doyle2014closed}, depend substantially on continuous data from CGMs and insulin pumps. A major barrier to progress in this field centers around access to rich datasets that facilitate the development of novel analytic solutions. In addition, there is a large amount of related but disconnected data streams that is not often reviewed or analyzed together, which further limits our understanding of diabetes management and even prevention \cite{walsh2015device,iyengar2016challenges}. To advance research and development of robust analytic solutions for the growing population of people with diabetes, there is a critical need for open datasets to understand outpatient management, develop interventions, and build clinically-relevant decision-support solutions.

\begin{figure}
    \centering
    \includegraphics[width=0.8\textwidth]{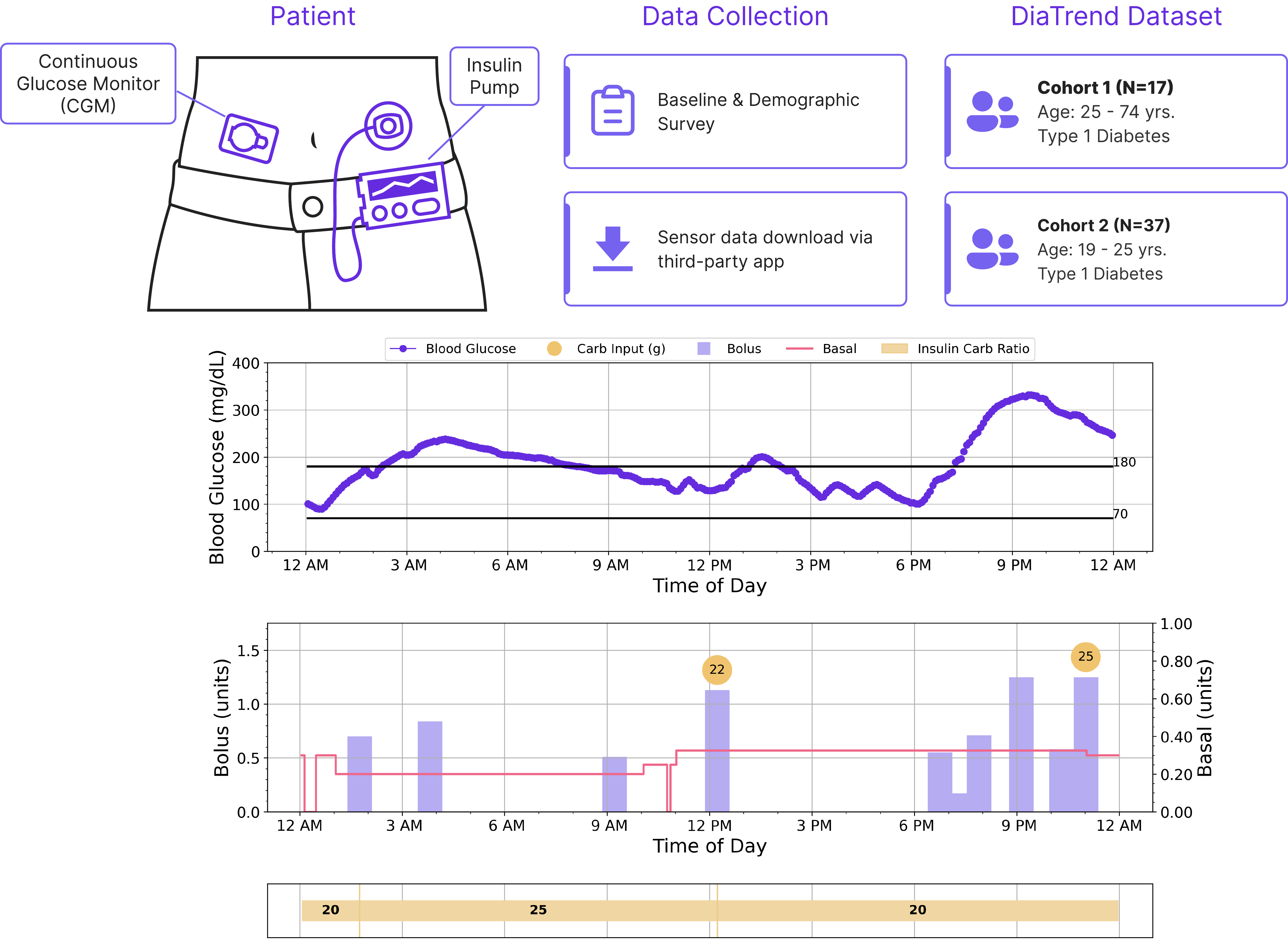}
    \caption{Overview of the data collection process and data provided in the DiaTrend dataset.}
    \label{fig:overview}
\end{figure}

Despite the recognized need for open datasets to enable research \cite{bietz2016opportunities}, there are limited datasets for data-driven research in the diabetes domain. One is the OhioT1DM dataset \cite{marling2020ohiot1dm}, which consists of eight weeks of CGM, insulin pump, physiological sensor, and self-reported events from 12 people with type 1 diabetes, while another is an N-of-1 dataset, which consists of two weeks of blood glucose, insulin, and carbohydrate intake logs \cite{Katz2018}. To broaden the scope of research on diabetes and chronic conditions in general, and accelerate development of robust computational solutions, we provide the DiaTrend dataset. The DiaTrend dataset includes CGM and insulin pump data from 54 patients with type 1 diabetes. This dataset is created from a subset of two larger studies focused on: 1) developing computational tools for self-management of diabetes \cite{bartolome2021glucomine}, and 2) evaluating a digital intervention for young adults with type 1 diabetes \cite{stanger2021digital}. The provided dataset includes time-aligned blood glucose samples recorded on average every 5 minutes with FDA-approved CGMs by Dexcom \cite{Dexcom_G6}, Abbott \cite{Abbott_FreeStyle}, and Medtronic \cite{Medtronic_Guardian}, and insulin pump data comprising basal and bolus insulin doses, carbohydrate intake logs, and other pump settings such as insulin-carb ratio and more. Figure \ref{fig:overview} presents an overview of the data collection process and data provided.

The DiaTrend dataset is useful for several research directions including more common tasks like blood glucose prediction \cite{gu2020neural,li2019convolutional,deng2021deep,zhu2018deep,li2019glunet,martinsson2018automatic,woldaregay2019data,zaidi2021multi}, prediction of adverse glycemic events (i.e., hypoglycemia and hyperglycemia) \cite{gadaleta2018prediction,mosquera2019leveraging,seo2019machine,dave2021feature}, detection of unannounced meals \cite{zheng2019automated,ramkissoon2018unannounced,xie2016variable,samadi2018automatic,kolle2019pattern}, and algorithm development for insulin delivery \cite{vettoretti2019combining,mosquera2023enabling}. However, this dataset is also useful to support further research on less studied topics like discovering digital biomarkers of glycemic control \cite{bartolome2022computational}, mining patterns/trends in diabetes management \cite{morton2020data,bartolome2021glucomine,belsare2022}, understanding adherence to wearable medical devices and patterns of missing data \cite{vhaduri2020adherence,drecogna2021data}, developing novel visual analytic and data visualization solutions \cite{zhang2018idmvis}, and designing decision-support tools through user-centered studies \cite{prioleau2020understanding,katz2018data,raj2019clinical,raj2019my}. Additionally, prospective researchers can find more opportunities for artificial intelligence in the diabetes domain through recent reviews in literature \cite{contreras2018artificial,ellahham2020artificial,tyler2020artificial}.

\section*{Methods}


\subsection*{Participants}
The DiaTrend dataset includes CGM and insulin pump data from a total of 54 patients with type 1 diabetes (age: 19 - 74 years, gender: 17 males, 37 females). Table \ref{Tab:subj_characteristics} provides an overview of the demographic and clinical characteristics of patients in this dataset, including the distribution across age groups, gender, race, diabetes type, and hemoglobin A1C. Participants were recruited through two independent studies. Study 1 (also known as Digital SMD) recruited patients from Dartmouth Health in 2019, while study 2 (also known as SweetGoals \cite{stanger2021digital}) is an ongoing randomized control trial that recruits patients through social media and online platforms. Both studies were approved by the Committee for Protection of Human Subjects at Dartmouth College and all participants provided verbal and written consent prior to joining either study. In addition, participants provided consent to share their data openly to the broader research community. 

Cohort 1 (n=17), from the Digital SMD study \cite{bartolome2021glucomine}, includes persons with type 1 diabetes between the ages of 25 to 74 years old who use a CGM and insulin pump for daily management of their condition and consented to share their retrospective device data for research. Meanwhile, cohort 2 (n=37), from the SweetGoals study \cite{stanger2021digital}, includes persons with type 1 diabetes for longer than 18 months between the ages of 19 to 29 years old who use a Glooko compatible glucometer or CGM, reported a clinical visit within the previous 6 months from the recruitment date, and self-reported their most recent Hemoglobin A1C (HA1C) value as >7.5\%. It is important to note that all device data included in the DiaTrend dataset was collected at baseline (i.e., prior to any intervention). Additionally, each individual's dataset spans varying time periods based on the available retrospective data at the time of recruitment. Given our focus on advanced diabetes technology for novel analytic solutions, only participants who use CGMs (with <30\% missing data) and insulin pumps for daily management are included in the provided dataset. 



\begin{table}
\centering
\begin{tabular}{l|cc}
\textbf{Characteristics}                       & \textbf{Count (n=54)} & \textbf{\%Dist}  \\
\hline \hline
\textbf{Age}                              &       &             \\
19 - 24 yrs                      & 24    & 44.4\%        \\
25 - 34 yrs                      & 14    & 25.9\%      \\
35 - 44 yrs                      & 4     & 7.4\%       \\
45 - 54 yrs                      & 3     & 5.6\%       \\
55 - 74 yrs                      & 9    & 16.7\%      \\

\hline
\textbf{Gender}                           &       &             \\
Female                           & 37    & 68.5\%      \\
Male                             & 17    & 31.5\%      \\

\hline
\textbf{Race}                             &       &             \\
White/Caucasian                  & 48    & 88.9\%      \\
Asian or Pacific Islander        & 2     & 3.7\%       \\
Black/African American           & 1     & 1.9\%       \\
Black/African American \& White & 1     & 1.9\%       \\
Other                           & 1    & 1.9\%      \\
Not Reported                     & 1    & 1.9\%      \\
\hline
\textbf{Diagnosis}                   &       &             \\
Type 1 Diabetes                 & 54    & 100\%      \\
\hline
\textbf{Hemoglobin A1C}                   &       &             \\
6.0-6.9                          & 8    & 14.8\%      \\
7.0-7.9                          & 23    & 42.6\%      \\
8.0-8.9                          & 17    & 31.5\%      \\
9.0-11.0                         & 3     & 5.6\%       \\
Not Reported                     & 3     & 5.6\%     
\end{tabular}
\caption{DiaTrend dataset: Demographic and clinical characteristics of patients with diabetes.}
\label{Tab:subj_characteristics}
\end{table}

\subsection*{Dataset Description}
\begin{figure}
    \centering
    \includegraphics[width=0.85\textwidth]{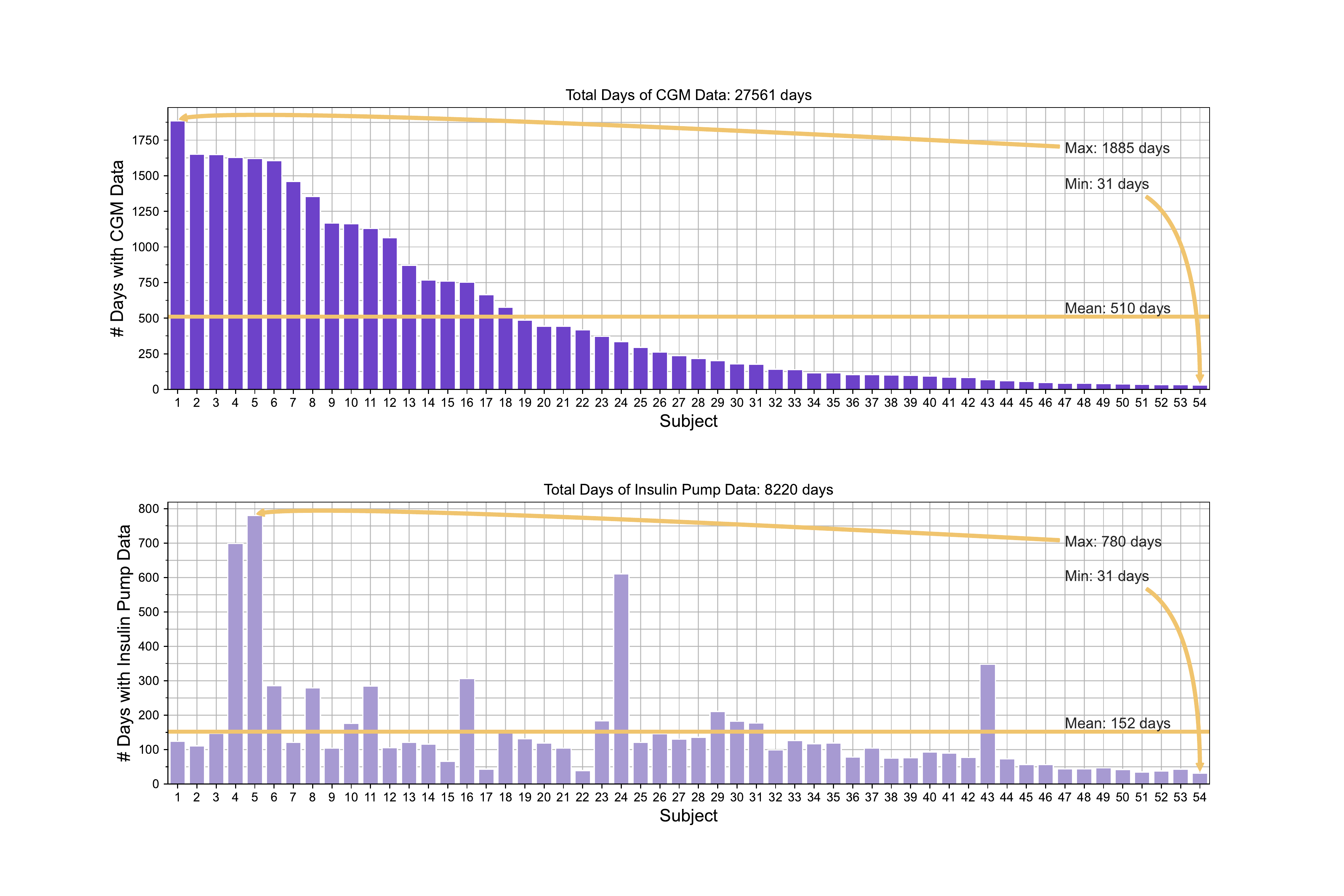}
    \caption{Overview of the number of days of sensor data per patient in the DiaTrend dataset.}
    \label{fig:sensor_data_duration}
\end{figure}

\begin{figure}
    \centering
    \includegraphics[width=0.85\textwidth]{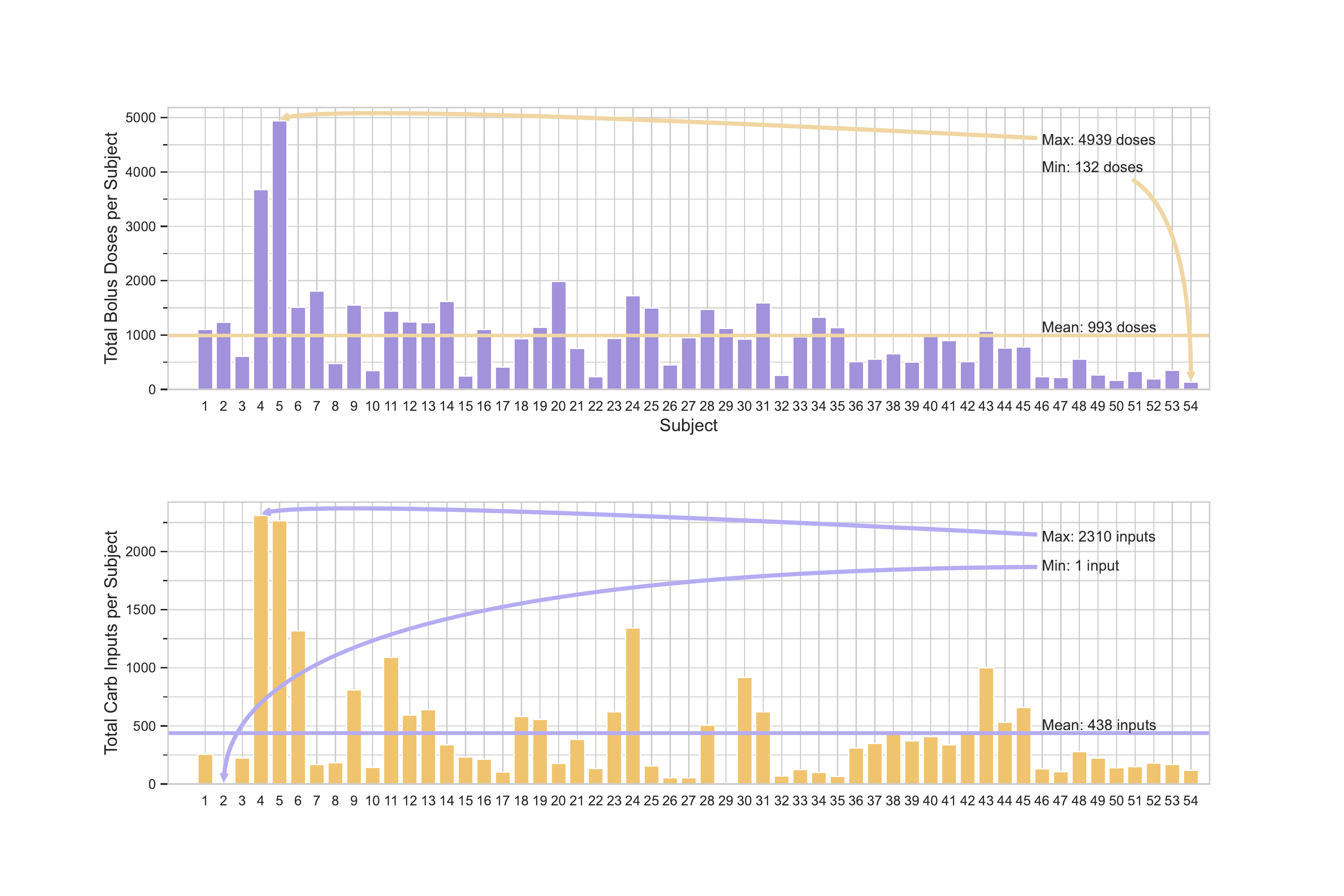}
    \caption{Overview of the total number of bolus and carb input data per patient in the DiaTrend dataset.}
    \label{fig:total_bolus_carb_input}
\end{figure}

The DiaTrend dataset includes a total of 27,561 days of CGM data and 8,220 days of insulin pump data from 54 patients with type 1 diabetes. In addition, the DiaTrend dataset includes demographic and clinical characteristics for each subject, including metrics such as age, gender, race, diabetes type and HA1C - see Table \ref{Tab:subj_characteristics}. There is an average of 510 days (range: 31 - 1885 days) of CGM data per subject, and an average of 152 days (range: 31 - 780 days) of insulin pump data per subject - see Fig. \ref{fig:sensor_data_duration}. Within the insulin pump data, there is an average of 993 total bolus doses per subject (range: 132 - 4939 doses) and an average of 438 total carb inputs per subject (range: 1 - 2310 input) - see Fig. \ref{fig:total_bolus_carb_input}. These data were collected as part of the Digital SMD \cite{bartolome2021glucomine} and SweetGoals \cite{stanger2021digital} studies during which each patient's retrospective CGM and insulin pump data was downloaded through a third-party application (i.e., Tidepool \cite{Tidepool} and Glooko \cite{Glooko}). It is important to note that since the SweetGoals study is a randomized control trial, only retrospective baseline data collected during the initial screening is included as part of the DiaTrend dataset (i.e., the provided data does not include sensor data from the intervention period of that study). In addition, HA1C - the primary clinically-validated metric for accessing glycemic control - was collected via the patient's electronic health record (i.e., the most recent HA1C) in the Digital SMD study and via a mail-in home test in the SweetGoals study at the time of the baseline assessment (approximately the endpoint of the device data).


%



\section*{Data Records}
All data records in the DiaTrend dataset are stored and accessible via the Synapse platform \cite{Prioleau2022_DiaTrend}. The deposited data consists of 54 Excel files--- one file for each subject. Each file has a CGM sheet that provides blood glucose data that was collected by the CGM. The CGM sheet includes 2 columns, namely, date and mg/dL. In addition, each subjects' file also has a Bolus sheet, which describes bolus insulin doses and meal announcements (i.e., user-entered estimates of carbohydrate content in meals logged to calculate bolus insulin needed to metabolize glucose from the meal consumed \cite{sora2019insulin}). The Bolus sheet includes the following 7 columns: date, normal, carbInput, insulinCarbRatio, bgInput, recommended.carb, and recommended.net. It is important to note that 37 (out of 54) Bolus sheets include 4 more columns, namely, recommended.correction, insulinSensitivityFactor, targetBloodGlucose, and insulinOnBoard. Additionally, there are only 17 subject files that have a Basal sheet, which describes the subject's basal infusions in 3 columns, namely, date, duration, and rate. Each row in all three of the Excel sheets refers to one record collected at a given timestamp in the column titled `date'. Excluding the date column, the rest of the data can be read as floating point numbers. Table \ref{Tab:data_records} provides a detailed breakdown of each data record, the format, and a description.

\begin{table}
\begin{tabular}{p{0.05\textwidth} p{0.2\textwidth} | p{0.3 \textwidth} p{0.35\textwidth}}
Sheet Name              & Column Name              & Format                       & Description  \\
\hline \hline
\multirow{2}{*}{CGM}    & date                     & datetime (yyyy-mm-dd HH:MM:SS) & {Date and time that glucose reading was recorded}\\ 
                        & mg/dL                    & Float64                      & Blood glucose reading in mg/dL \\
\hline
\multirow{11}{*}{Bolus} & date                     & datetime (yyyy-mm-dd HH:MM:SS) & Date and time that bolus was administered \\ 
                        & normal                   & Float64                      & Amount of bolus insulin delivered (units) \\ 
                        & carbInput                & Float64                      & Total carbs announced for bolus (grams) \\
                        & insulinCarbRatio         & Float64                      & Patient setting for grams of carbs covered per one unit of insulin \\ 
                        & bgInput                  & Float64                      & Blood glucose reading at time of bolus (mg/dL) \\ 
                        & recommended.carb         & Float64                      & Amount of insulin recommended to cover carb intake for normal bolus \\ 
                        & recommended.net          & Float64                      & Amount of insulin recommended for bolus delivery \\ 
                        & recommended.correction   & Float64                      & Amount of insulin recommended for correction component of normal bolus \\ 
                        & insulinSensitivityFactor & Float64                      & Patient setting for how one unit of insulin lowers blood glucose level \\ 
                        & targetBloodGlucose       & Float64                      & Target blood glucose value for after bolus delivery  \\ 
                        & insulinOnBoard           & Float64                      & Amount of active insulin remaining from prior insulin doses \\
\hline
\multirow{2}{*}{Basal}  & date                    & datetime (yyyy-mm-dd HH:MM:SS) & Date and time of basal infusion\\ 
                        & duration                & Float64                      &  Duration of basal infusion (ms)\\   
                        & rate                    & Float64                      &  Rate of basal infusion (units/hr) 

\end{tabular}
\caption{Overview of the data records, format, and description in the DiaTrend dataset.}
\label{Tab:data_records}
\end{table}



\section*{Technical Validation}
\begin{figure}
    \centering
    \begin{subfigure}{0.95\textwidth}
         \includegraphics[width=\textwidth]{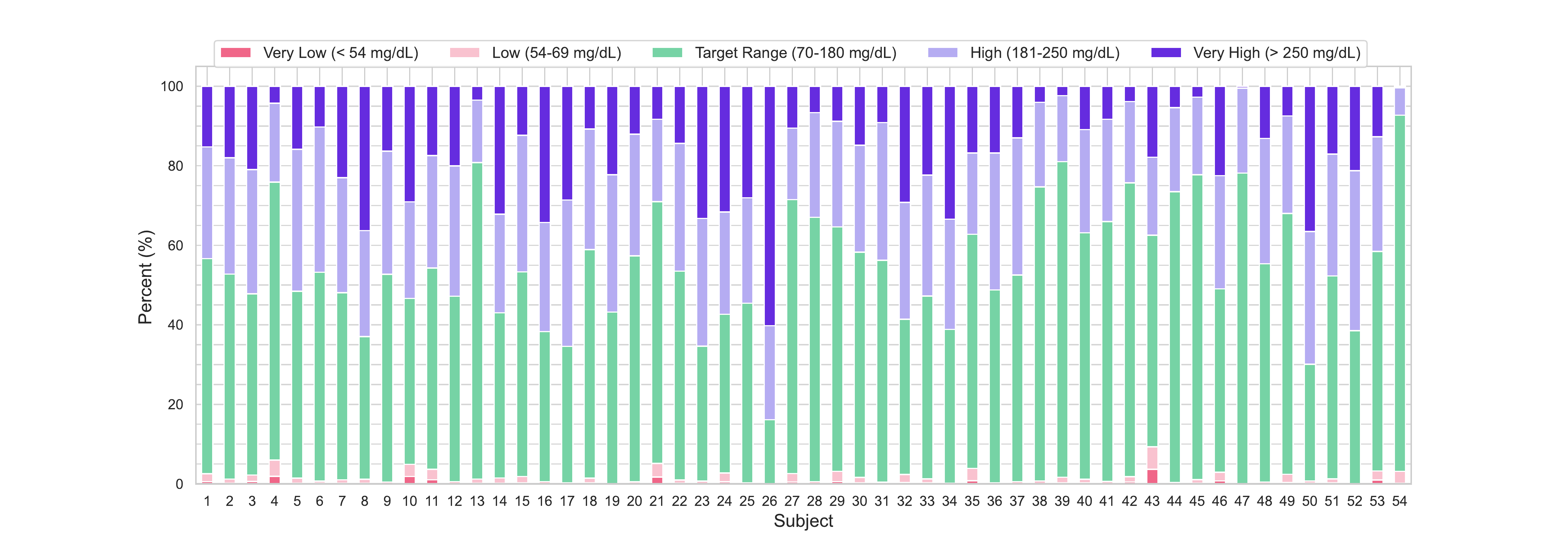}
         \caption{}
    \label{fig:time-in-ranges}
    \end{subfigure}
    \begin{subfigure}{0.95\textwidth}
        \includegraphics[width=\textwidth]{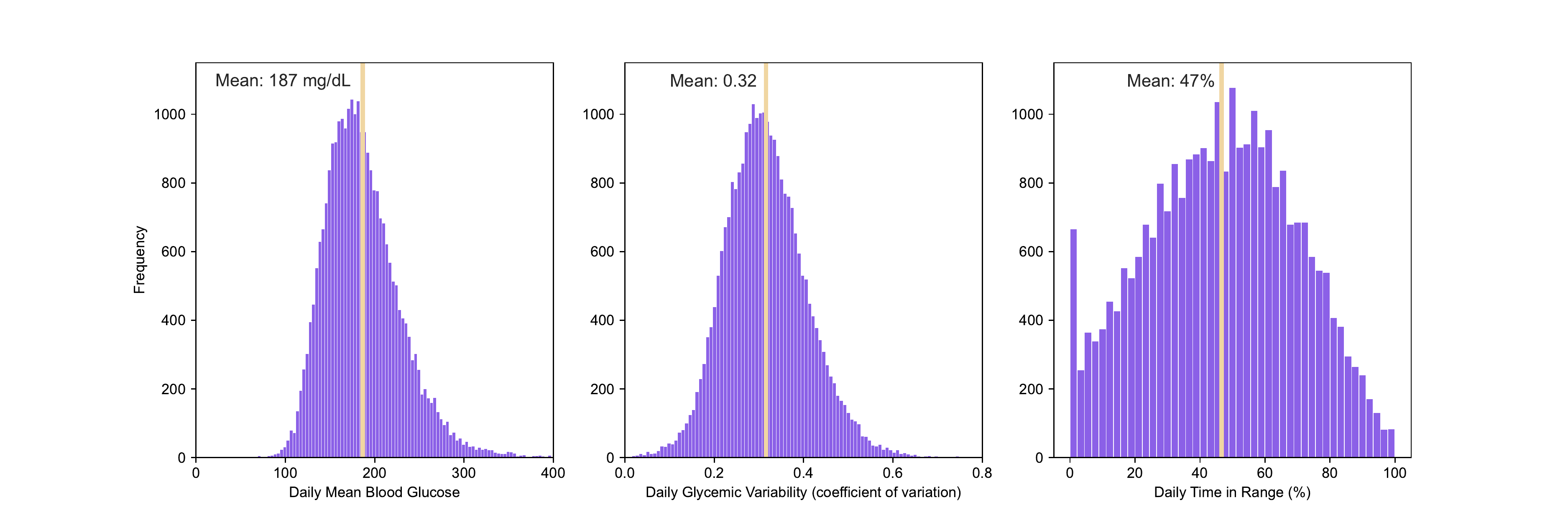}
        \caption{}
        \label{fig:cgm_daily_histogram}
    \end{subfigure}
    \caption{Descriptive summary of CGM data in the DiaTrend dataset. (a) The percent of blood glucose samples in 5 clinically-relevant categories. (b) The distributions of daily mean blood glucose, daily glycemic variability, and daily time in [target] range.}
\end{figure}

For each patient included in the DiaTrend dataset, we provide an overview of their blood glucose data using clinically-validated metrics for assessing glycemic control \cite{battelino2019clinical,danne2017international}. This includes the percentage of all blood glucose readings in 5 clinically-relevant categories, namely, very low (< 54 mg/dL), low (54 - 69 mg/dL), target range (70 - 180 mg/dL), high (181 - 250 mg/dL), and very high (> 250 mg/dL). From Fig. \ref{fig:time-in-ranges}, we can observe that blood glucose is highly variable and only a minority of patients living with diabetes (less than 10\% in our dataset) meet the clinical target of maintaining blood glucose within the target range of 70 - 180 mg/dL for more than 70\% of the time \cite{battelino2019clinical}. Fig. \ref{fig:cgm_daily_histogram} presents histograms for daily mean blood glucose (mean = 187 mg/dL), daily glycemic variability (mean = 0.33), and daily time in range (mean = 47\%). From this figure, we can observe a normal distribution for each clinically-relevant metric in the DiaTrend dataset. 

\begin{figure}
     \centering
    \begin{subfigure}{0.95\textwidth}
        \includegraphics[width=0.95\textwidth]{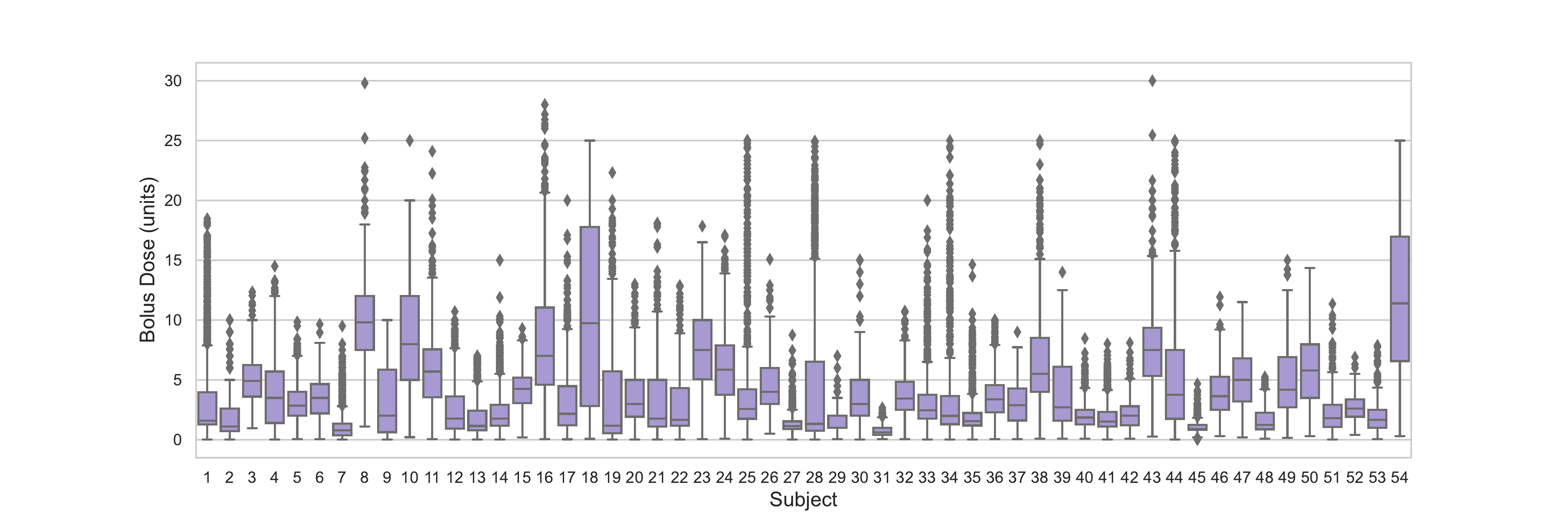}
    \caption{}
    \label{fig:bolus_insulin_boxplot}
    \end{subfigure}
    \centering
    \begin{subfigure}{0.95\textwidth}
        \includegraphics[width=0.95\textwidth]{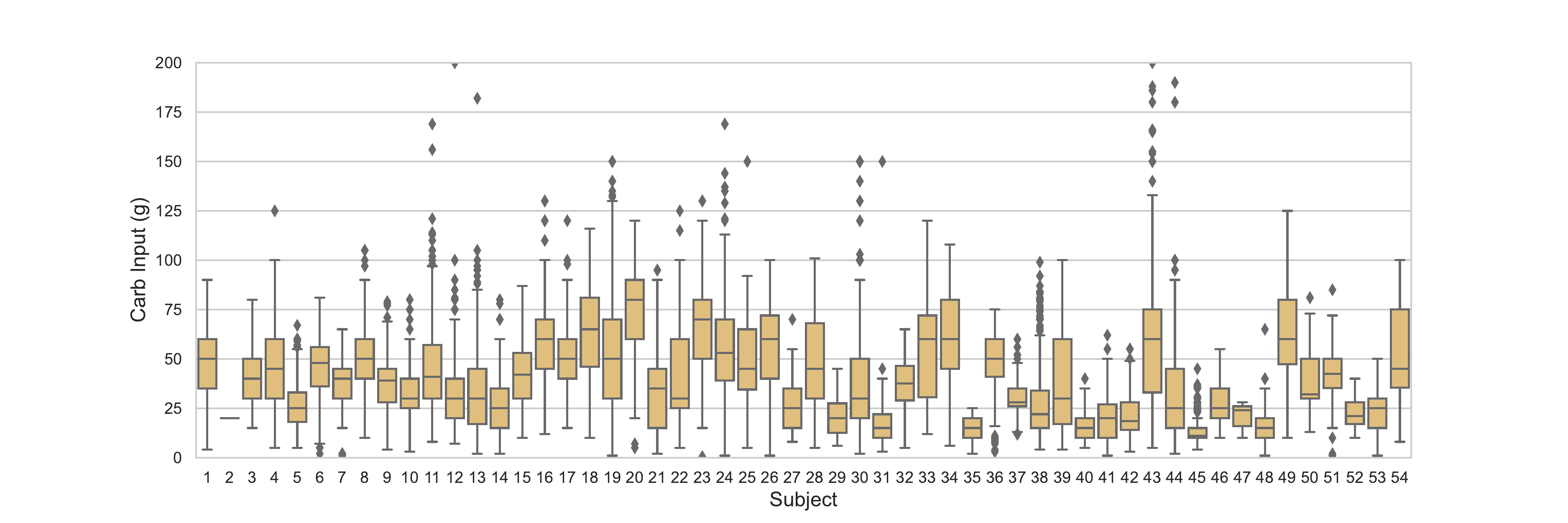}
    \caption{}
    \label{fig:carbinput_boxplot}
    \end{subfigure}
    \begin{subfigure}{0.95\textwidth}
        \centering
        \includegraphics[width=0.95\textwidth]{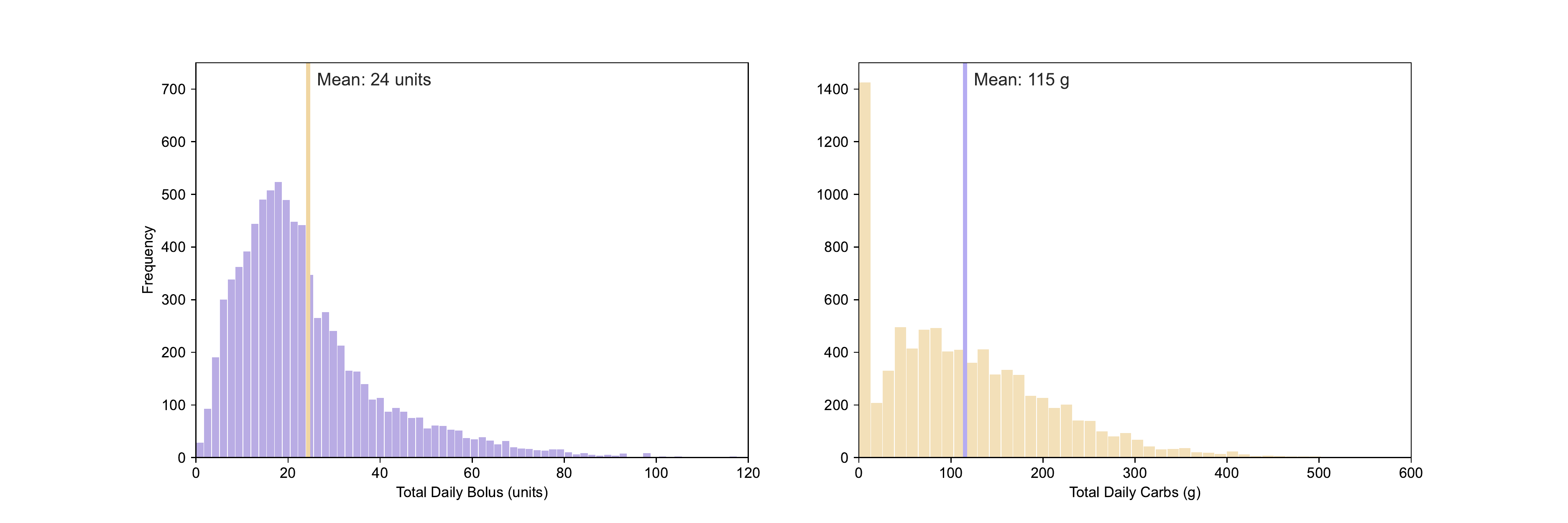}
        \caption{}
        \label{fig:insulinpump_dailyhist}
    \end{subfigure}
    \caption{Descriptive summary of insulin pump data in the DiaTrend dataset. (a) A box plot of all bolus insulin doses per subject. (b) A box plot of all carb input entries per subject. (c) The distributions of total daily bolus insulin and total daily carb inputs across all subjects. }
        \label{fig:overview_insulinpump_data}
\end{figure}

Similarly, we provide an overview of each patient's insulin pump data using box plots and histograms. Fig. \ref{fig:bolus_insulin_boxplot} and \ref{fig:carbinput_boxplot} show box plots with descriptive statistics associated with bolus insulin doses and carb inputs, respectively, for each subject. Additionally, Fig. \ref{fig:insulinpump_dailyhist} shows the distributions of total daily bolus insulin doses (units) and total daily carb inputs (g), respectively. From this figure, we can observe a mean total daily bolus of 24 units and a mean total daily carb input of 115 g, both with a positively skewed distribution. In particular, we observe a high number of days ($\sim$1400 days) with very low carb inputs ($\sim$0g); this could be indicative of missed mealtime boluses (i.e., no bolus insulin used during mealtimes) --- this is a common contributor to poor glycemic outcomes \cite{burdick2004missed,deeb2015important,patton2014frequency}.

\subsection*{Limitations}
There are some important considerations and limitations associated with the DiaTrend dataset provided in this paper. First, there is imbalance in the representation of subjects across the dimensions of race, gender, and age. More specifically, majority of patients whose CGM and insulin pump data is provided (i.e., 48 out of 54 or 89\%) are non-Hispanic White/Caucasian. Also, this dataset includes a lower representation of males (n=17 out of 54 or 32\%) compared to females, and a lower representation of older adults (e.g., for age $\geq$ 45 years old, n=12 or 22\%). The limitation with regards to race (i.e., low representation of participants from non-White/Caucasian races including Asian and Black/African Americans) is partly due to the geographical location (i.e., New Hampshire) from which some participants were recruited. However, the imbalance in representation also underscores racial disparities which have been identified in prior literature relating to access and use of advanced diabetes technologies \cite{foster2019state}, particularly CGMs and insulin pumps. Additionally, the limitation with regards to age (i.e., low representation of older adults and higher representation of young adults) is primarily due to the targeted focus on young adults with type 1 diabetes in the SweetGoals study \cite{stanger2021digital}. A second limitation of the DiaTrend dataset is that it lacks full temporal alignment in the CGM and insulin pump data for each participant. This difference is apparent from Fig. \ref{fig:sensor_data_duration} which shows more CGM data than insulin pump data for a number of subjects. While the reason for this is unknown, we suspect that it is primarily due lower data storage capacity on insulin pumps compared to CGMs, which in turn limits the amount of retrospective data available for download from insulin pumps. Third, basal insulin data is not available for subjects from cohort 2 (37 out of 54). This missing data stream might limit research efforts that require basal rate for analysis. However, despite the aforementioned limitations, the DiaTrend dataset represents one of the largest open-source datasets currently available in the diabetes domain. This critical resource provides a unique opportunity to advance development of novel data-driven solutions that can improve the lives of people living with diabetes. In addition, this dataset provides a necessary benchmark to evaluate the generalizability of numerous diabetes-relevant algorithms in literature \cite{gu2020neural,li2019convolutional,deng2021deep,zhu2018deep,li2019glunet,woldaregay2019data,zaidi2021multi,gadaleta2018prediction,mosquera2019leveraging,seo2019machine,dave2021feature,zheng2019automated,ramkissoon2018unannounced,xie2016variable,samadi2018automatic,martinsson2018automatic,vettoretti2019combining,kolle2019pattern}. 

\section*{Usage Notes}
The DiaTrend dataset is provided for research and educational purposes that support the development of novel data-driven solutions for the diabetes community and beyond. Consistent with exemplar studies \cite{marling2020ohiot1dm,bot2016mpower,hershman2019physical}, we have set governance structures in place to balance the need for open datasets that advance research and protect the privacy of participants.

Researchers interested in accessing the DiaTrend dataset should complete the following steps:
\begin{enumerate}
    \item Register for a Synapse account (www.synapse.org).
    \item Become a Synapse Certified User with a validated user profile.
    \item Submit an Intended Data Use statement.
    \item Agree to the Conditions of Use.
\end{enumerate}

The conditions of use are as follows:
\begin{itemize}
    \item You confirm that you will not attempt to re-identify research participants for any reason, including for re-identification theory research.
    \item You commit to keeping the DiaTrend dataset confidential and secure.
    \item You understand that these data may not be used for commercial advertisement or to re-contact research participants.
    \item You agree to acknowledge the research participants as data contributors, study investigators, and this paper on all publications or presentations which results from using the DiaTrend dataset.
\end{itemize}



\section*{Code availability}
Python was used for all data processing described in this paper. The Python code used to generate all figures in this paper is available on the Augmented Health Lab's Github: https://github.com/Augmented-Health-Lab/Diatrend.

\bibliography{references}


\section*{Acknowledgements}
C.S. acknowledges funding from the National Institute of Diabetes and Digestive and Kidney Diseases (R01DK124428).

\section*{Author contributions statement}

T.P. conceived the Digital SMD study for cohort 1. T.P, A.B., and R.C. led efforts for data collection from cohort 1. C.S. conceived and led the SweetGoals study and data collection for cohort 2. A.B. cleaned and organized both datasets. T.P. and A.B. wrote the manuscript, and prepared figures \& tables. All authors reviewed and contributed to the final approval of the manuscript. 


\section*{Competing interests}

The authors declare no completing interest.

\end{document}